\documentclass[aps,pra,twocolumn,longbibliography] {revtex4-1}
\usepackage{amsmath,graphicx,amsfonts}

\usepackage{times}
\usepackage[colorlinks=true,linkcolor=blue,urlcolor=blue,citecolor=blue, breaklinks=true,hypertexnames=false]{hyperref}

\usepackage{amssymb}
\usepackage{bbm}

\usepackage[utf8]{inputenc}
\DeclareUnicodeCharacter{03BC}{\mbox{$\mu$}}

\DeclareUnicodeCharacter{03B4}{$\delta$}
\DeclareUnicodeCharacter{2009}{-}

\usepackage[english]{babel}
\addto\captionsenglish{}

\makeatletter
\def\bbl@set@language#1{%
	\edef\languagename{%
		\ifnum\escapechar=\expandafter`\string#1\@empty
		\else\string#1\@empty\fi}%
	\@ifundefined{babel@language@alias@\languagename}{}{%
		\edef\languagename{\@nameuse{babel@language@alias@\languagename}}%
	}%
	\select@language{\languagename}%
	\expandafter\ifx\csname date\languagename\endcsname\relax\else
	\if@filesw
	\protected@write\@auxout{}{\string\select@language{\languagename}}%
	\bbl@for\bbl@tempa\BabelContentsFiles{%
		\addtocontents{\bbl@tempa}{\xstring\select@language{\languagename}}}%
	\bbl@usehooks{write}{}%
	\fi
	\fi}
\newcommand{\DeclareLanguageAlias}[2]{%
	\global\@namedef{babel@language@alias@#1}{#2}%
}
\makeatother

\DeclareLanguageAlias{en}{english}

\makeatletter
\def\@bibdataout@aps{%
	\immediate\write\@bibdataout{%
		@CONTROL{%
			apsrev41Control%
			\longbibliography@sw{%
				,author="08",editor="1",pages="1",title="0",year="1"%
			}{%
				,author="08",editor="1",pages="1",title="",year="1"%
			}%
		}%
	}%
	\if@filesw \immediate \write \@auxout {\string \citation {apsrev41Control}}\fi 
}
\makeatother

\date\today
\begin{document}
\title{Dispersion relation of a polaron in the Yang-Gaudin Bose gas}
\author{Zoran Ristivojevic}
\affiliation{Laboratoire de Physique Th\'{e}orique, Universit\'{e} de Toulouse, CNRS, UPS, 31062 Toulouse, France}
	
\begin{abstract}
We study a one-dimensional Bose gas with two internal states described by the Yang-Gaudin model and calculate analytically the dispersion relation of a polaron quasiparticle, which is the lowest excitation branch. We study the dispersion in the thermodynamic limit in the regimes of weak and strong interaction without limitations on the momentum. At weak interaction the polaron dispersion is in the vicinity of the dark soliton one; we calculate the leading deviation in the parametric form. At strong interaction we find an ansatz for the explicit form of the polaron dispersion. It has the form of a power series of the sine function of the momentum with interaction-dependent coefficients. By increasing the power of the series, the corresponding coefficients show faster decay and thus one practically needs only a few of them; we give the results for the first three. The coefficients of the series are connected to the Maclaurin series of the polaron dispersion and thus it is sufficient to calculate latter quantity to automatically find the power series result for the dispersion at all momenta. The derived results for the dispersion can be used to obtain explicit expressions for the exponents of the power-law singularities in the response functions at the spectral edge.
\end{abstract}
\maketitle

\section{Introduction}

A conducting electron in polar crystals produces the distortions of the ion lattice due to Coulomb interaction. In turn, the resulting quanta of lattice vibrations, phonons, act on the electron changing its properties. The electron moves together with an accompanying cloud of phonons. The resulting composite quasiparticle is known as a polaron \cite{appel_polarons_1968}. Its own characteristics are the effective mass, momentum, and energy. Due the interaction with phonons, the effective polaron mass is larger than the one of the electron, while its energy is smaller \cite{landau_effective_1948}.

Once developed for electrons in crystals, the whole concept has become broader. The polaron problem is a paradigm for a system of a quantum impurity interacting with the many-body environment. In this context, the case of a one-dimensional host is particularly interesting due to the existence of an exact  theoretical method to treat the problem in certain situations: the Bethe ansatz. This technique results in a set of algebraic equations that hide the relevant physics. Their analysis is, however, typically an involved problem. 

A well-known Hamiltonian for the polaron problem that is solved by the Bethe ansatz is the one of spin-$\frac{1}{2}$ fermions in one dimension, where 
the noninteracting fermions of the same spin interact via a local delta-function potential with a single fermion of the opposite spin. Both cases of repulsive and attractive interaction are considered and explicit analytical results for the ground and excited states  are obtained \cite{mcguire_interacting_1965,mcguire_interacting_1966}. A bosonic counterpart of this system is also exactly solved \cite{yang_exact_1967,gaudin_2014,li_exact_2003,oelkers_bethe_2006}. It consists of one-dimensional bosons with two internal degrees of freedom, which are formally described by isospin-$\frac{1}{2}$ bosons. In the present case the same local repulsion is assumed among all the bosons. The Hamiltonian for both problems is known under the name Yang-Gaudin. 

In this paper we study the Yang-Gaudin Hamiltonian for the Bose gas. For the purpose of discussing the spectrum of elementary excitations, it is sufficient to consider the bath of bosons all of the same isospin with one boson of the opposite isospin. This system has three types of elementary excitations, where two of them are the ones of the single-component Lieb-Liniger Bose gas \cite{lieb_exact_1963,lieb_exact2_1963}. They are classified as particlelike type-I and holelike type-II excitations. In the regime of weak repulsion and away from the tiny range of very small momenta \cite{imambekov_one-dimensional_2012,pustilnik_low-energy_2014,ristivojevic_excitation_2014}, the excitations can be described semiclassically 
\cite{kulish_comparison_1976}. They are characterized by the Bogoliubov spectrum and the spectrum of dark solitons, respectively \cite{ishikawa_solitons_1980}. The third kind of excitation arises due to the presence of an extra boson with the opposite isospin and is known under the names spin-wave excitation or magnon \cite{fuchs_spin_2005,zvonarev_edge_2009,matveev_spectral_2008,kamenev_dynamics_2009} and isospinon \cite{li_exact_2003}. In this paper we find it appropriate to call it polaron quasiparticle excitation, which is represented by this collective excitation, and speak about its dispersion. This is the focus of the paper. Numerical studies of the polaron dispersion are performed for systems of a finite size  \cite{li_exact_2003,robinson_excitations_2017} as well as in the thermodynamic limit \cite{zvonarev_edge_2009}. On the other hand, explicit analytical results are obtained in the regimes of small momenta and in the limiting cases of interaction \cite{fuchs_spin_2005,batchelor_collective_2006,matveev_spectral_2008,ristivojevic_exact_2021}.

At low momenta, the dispersion relation of a polaron in a one-dimensional Bose liquid is quadratic \cite{fuchs_spin_2005,zvonarev_edge_2009,matveev_spectral_2008,kamenev_dynamics_2009,lamacraft_dispersion_2009,schecter_critical_2012,Note1}\footnotetext[1]{The latter is correct beyond one dimension \cite{halperin_dynamic_1975}}.\nocite{halperin_dynamic_1975} Since the lowest excitations of the liquid are phonons with linear dispersion, it is energetically more favorable for the system to host an excited polaron than a phonon. This picture is valid beyond the small-momentum regime as the polaron dispersion lies below the type-II excitation branch \cite{zvonarev_edge_2009}. Thus the polaron branch is the lowest excitation mode in the system. In the thermodynamic limit, the polaron dispersion is a periodic function, $\mathcal{E}(p)=\mathcal{E}(p+2\pi\hbar n)$, where $p$ is the polaron momentum, and $n$ is the density of the liquid \cite{kamenev_dynamics_2009,lamacraft_dispersion_2009,schecter_critical_2012}. This property is discussed later in the text, as well as the reflection property of the dispersion around the momentum $\pi\hbar n$, which is $\mathcal{E}(\pi\hbar n+q)=\mathcal{E}(\pi\hbar n-q)$. The two properties enable us to study the polaron dispersion in the limited region $0< p\le \pi\hbar n$, since they automatically determine the dispersion at other momenta.

In the reminder of the paper, in Sec.~\ref{sec1} we introduce the Yang-Gaudin Hamiltonian for the Bose gas and briefly review its Bethe ansatz solution relevant for a single polaron. We discuss the limit corresponding to the Lieb-Liniger model as well as the parametric form of the polaron dispersion branch, showing some of its general properties. Section~\ref{sec3} contains a study of the dispersion at weak interaction and Sec.~\ref{sec4} contains a study of the dispersion at strong interaction, without limitation on the momenta. We found an ansatz in terms of a series that describes the dispersion at strong interaction, which appears to be asymptotically exact, and we calculated the leading-order terms. In Sec.~\ref{sec5} an elementary calculation of the polaron dispersion at weak interaction and low momenta is discussed. In the Appendix, we present an efficient, highly-precise numerical procedure to calculate the polaron dispersion, which is used to produce the plots.

\section{Yang-Gaudin model for the Bose gas\label{sec1}} 

A one-dimensional single-component Bose gas with contact repulsion is described by the Hamiltonian
\begin{align}\label{eq:H}
H=\frac{\hbar^2}{2m}\left[-\sum_{j=1}^{N} \frac{\partial^2}{\partial x_j^2}+c\sum_{j\neq l} \delta(x_j-x_l)\right],
\end{align}
which was studied by Lieb and Liniger \cite{lieb_exact_1963}.
In Eq.~(\ref{eq:H}), $m$ is the particle mass, $c$ controls the interaction strength, and $N$ is the total number of particles. For simplicity we assume periodic boundary conditions in the following. The Hamiltonian (\ref{eq:H}) is remarkable as it admits an exact solution, which is obtained by the Bethe ansatz technique \cite{lieb_exact_1963,lieb_exact2_1963}.

Interestingly, the Hamiltonian (\ref{eq:H}) is also solvable for a two-component (i.e., isospin-$\frac{1}{2}$) Bose gas \cite{yang_exact_1967,gaudin_2014,li_exact_2003,oelkers_bethe_2006}. In this case Eq.~(\ref{eq:H}) is known as the Yang-Gaudin model. Its eigenstates can be classified with respect to the value of the total isospin. In the sector where it has the maximal value $N/2$, the system simplifies to the single-component Lieb-Liniger model. It is characterized by $N$ density quantum numbers $I_1,I_2,\ldots, I_N$ that define $N$ quasimomenta, $k_1,k_2,\ldots, k_N$. In the case of the total isospin $N/2-1$, which is the focus of this paper, the system acquires an additional spin quantum number $J$ that defines the spin rapidity $\eta$. The Bethe ansatz equations for the Yang-Gaudin model of the Bose gas are given by \cite{li_exact_2003,oelkers_bethe_2006}
\begin{subequations}\label{BA}
\begin{gather}
	e^{i k_j L}=-\frac{k_j-\eta-i c/2}{k_j-\eta+i c/2}\prod_{l=1}^{N}\frac{k_j-k_l+i c}{k_j-k_l-i c},\\
	1=\prod_{l=1}^{N}\frac{\eta-k_l-i c/2}{\eta-k_l+i c/2},
\end{gather}
\end{subequations}
where an integer $j$ satisfies $1\le j\le N$ and $L$ is the system size. Using $\ln \frac{A+ic}{A-ic}=2i \arctan\frac{c}{A}$ and $\arctan x+\arctan(1/x)=\pi\mathrm{sgn}(x)/2$, the system (\ref{BA}) can be brought to the following form \cite{zvonarev_edge_2009}:
\begin{subequations}\label{eq:BE}
\begin{gather}\label{eq:BE1}
L k_j+\sum_{l=1}^N \theta(k_j-k_l)={2\pi}I_j+\pi+\theta(2k_j-2\eta),\\\label{eq:BE2}
2\pi J=\sum_{l=1}^N \theta(2\eta-2k_l).
\end{gather} 
\end{subequations}
Here $I_j=n_j-(N+1)/2$, where $n_j$ are integers, $J$ is an integer or odd half-integer depending on whether $N$ is even or odd, and
\begin{align}\label{eq:phaseshift}
\theta(k)=2\arctan(k/c).
\end{align}
Equation (\ref{eq:phaseshift}), up to the sign, denotes the scattering phase shift. It satisfies $\theta(\pm \infty)=\pm \pi$ and $\theta(-k)=-\theta(k)$. The energy $E$ and the momentum $p$ of the system are given by
\begin{align}\label{eq:EP}
E=\frac{\hbar^2}{2m}\sum_{j=1}^N k_j^2,\quad p=\hbar \sum_{j=1}^N k_j.
\end{align}
Note that the spin rapidity $\eta$ indirectly enters Eq.~(\ref{eq:EP}) through the Bethe ansatz equations.

\subsection{Limit of the Lieb-Liniger model}

In the special case $\eta\to+\infty$, Eq.~(\ref{eq:BE1}) becomes independent of the spin quantum number $J$ that enters Eq.~(\ref{eq:BE2}), and describes the quasimomenta of the single-component system of bosons described by the Lieb-Liniger model \cite{lieb_exact_1963}. Its ground state is realized for
\begin{align}\label{eq:Ij}
I_j=j-\frac{N+1}{2},\quad j=1,2,\ldots,N.
\end{align}
Instead of studying the quasimomenta in Eq.~(\ref{eq:BE1}), it is useful to introduce the density of quasimomenta via the relation $\rho(k_j)=[L(k_{j+1}-k_j)]^{-1}$. In the thermodynamic limit, $L,N\to\infty$ such that the density $n=N/L$ is kept constant, it satisfies the integral equation \cite{lieb_exact_1963}
\begin{align}\label{eq:rho}
\rho(k,Q)-\frac{1}{2\pi}\int_{-Q}^{Q} dq\:\!\theta'(k-q)\rho(q,Q)=\frac{1}{2\pi}.
\end{align}
Here $Q$ denotes the Fermi rapidity, which is the largest quasimomentum in the ground state. The kernel in the integral operator is determined by the phase shift (\ref{eq:phaseshift}), $\theta'(k)=d\theta(k)/dk$.

The density of quasimomenta in the ground state is a symmetric (even) function. It enables us to calculate the density, the momentum, and the energy of the system. The density of particles is given by
\begin{align}\label{eq:n}
n(Q)=\int_{-Q}^{Q} dk\:\!\rho(k,Q),
\end{align}
the ground-state momentum is zero, and the ground-state energy per particle $\epsilon_0$ can be expressed in the form \cite{lieb_exact_1963}
\begin{align}\label{eq:E0}
\epsilon_0=\frac{\hbar^2}{2m\:\!n}\int_{-Q}^{Q} dk\:\!k^2\rho(k,Q).
\end{align}
We note that the Fermi rapidity $Q$ naturally enters $\epsilon_0$ in Eq.~(\ref{eq:E0}) and $n$ in Eq.~(\ref{eq:n}). However, if needed, one can express $Q$ in terms of $n$ using their connection (\ref{eq:n}).

\subsection{Momentum and energy of the polaron excitation}

As follows from Eq.~(\ref{eq:BE2}), the spin quantum number takes the maximal value $J=N/2$ at $\eta\to+\infty$. Let us study the case of finite $\eta$ where $I_j$ assumes the values given by Eq.~(\ref{eq:Ij}). Then Bethe ansatz Eqs.~(\ref{eq:BE}) describe the excited state of the system that hosts a magnon \cite{fuchs_spin_2005,zvonarev_edge_2009}, which in this paper  represents a polaron quasiparticle excitation. The momentum of the system in the excited state coincides with the momentum of the polaron excitation. It can be obtained from Eqs.~(\ref{eq:EP}) and (\ref{eq:BE1}), leading to $p=(\hbar/L)\sum_{j=1}^{N}\left[\pi+\theta(2k_j-2\eta)\right]$. In the thermodynamic limit this gives 
\begin{subequations}
	\label{eq:peps}
\begin{align}\label{eq:p}
p(Q,\eta)=\hbar \int_{-Q}^{Q} dk\:\! \rho(k,Q)\left[\pi+\theta(2k-2\eta)\right].
\end{align}
Notice that $p$ explicitly depends on the Fermi rapidity $Q$ and the spin rapidity $\eta$. 

Evaluation of the energy of the system in the excited state from Eq.~(\ref{eq:EP}) is more involved \cite{ristivojevic_exact_2021}. The final result in the thermodynamic limit can be expressed as $E=N\epsilon_0+\mathcal{E}(Q,\eta)$, where $\epsilon_0$ is given by Eq.~(\ref{eq:E0}), while $\mathcal{E}(Q,\eta)$ is the energy of polaron excitation corresponding to the momentum (\ref{eq:p}). It is given by
\begin{align}\label{eq:eps}
\mathcal{E}(Q,\eta)=\frac{1}{2\pi}	\int_{-Q}^{Q} dk\:\! \sigma(k,Q)\theta(2k-2\eta).
\end{align}
\end{subequations}
Here $\sigma(k,Q)$ satisfies the integral equation
\begin{align}\label{eq:sigma}
	\sigma(k,Q)-\frac{1}{2\pi}\int_{-Q}^{Q} dq\:\!\theta'(k-q)\sigma(q,Q)=\frac{\hbar^2}{m}k,
\end{align}
where the kernel is determined by Eq.~(\ref{eq:phaseshift}). 

Equations  (\ref{eq:peps}) are exact and determine the dispersion of the polaron excitation in the parametric form. Together with Eqs.~(\ref{eq:rho}),  (\ref{eq:sigma}), and (\ref{eq:n}), we have obtained a closed set that should be eventually inverted in order to find the explicit form of the dispersion $\mathcal{E}(p)$. Analytically, this is a difficult problem for arbitrary values of the interaction. Here we study the regimes of weak and strong interactions without restriction on the momentum $p$. 

\subsection{General properties of the polaron spectrum}

Before evaluating the dispersion, let us first reveal its global features. The momentum and energy (\ref{eq:peps}) satisfy some general properties that follow from the general properties of $\rho(k,Q)$ and $\sigma(k,Q)$ that are, respectively, even and odd analytic functions of $k$. It then follows that the momentum and energy (\ref{eq:peps}) are analytic functions of $\eta$.  The momentum (\ref{eq:p}) is bounded between $p(Q,\eta\to+\infty)=0$ and $p(Q,\eta\to-\infty)=2\pi\hbar n$. It has the symmetry property around $\pi\hbar n$,
\begin{align}\label{eq:psim}
	p(Q,\eta)-\pi\hbar n=\pi\hbar n-p(Q,-\eta),
\end{align}
which implies $p(Q,0)=\pi\hbar n$. 

The energy (\ref{eq:eps}) is an even function of $\eta$. It is bounded and reaches the minimum at $\mathcal{E}(Q,\eta\to\pm\infty)=0$, while the maximum occurs at $\eta=0$, which is
\begin{align}
	\mathcal{E}(Q,0)=\frac{1}{\pi}\int_{0}^{Q}dk \sigma(k,Q)\theta(2k).
\end{align} 
The energy at the maximum thus satisfies the inequality
\begin{align}\label{eq:ineq}
	\mathcal{E}(Q,0)<\int_{0}^{Q}dk \sigma(k,Q).
\end{align}
The quantity on the right-hand side of Eq.~(\ref{eq:ineq}) formally denotes the energy of the type-II excitation with the momentum $\pi\hbar n$ in the Lieb-Liniger model \cite{lieb_exact2_1963,reichert_exact_2019}. Therefore, the energy of the polaron excitation at its maximum is smaller than the energy of the type-II excitation. This result is valid at any repulsion as numerically verified in Ref.~\cite{zvonarev_edge_2009}.

The parity of the energy (\ref{eq:eps}) with respect to $\eta$, in conjunction with Eq.~(\ref{eq:psim}), gives the symmetry property of the dispersion when it is expressed explicitly as a function of the momentum \cite{Note2}\footnotetext[2]{Note that from this moment the notation becomes ambiguous since $\mathcal{E}$ was initially defined as a function of two arguments, see  Eq.~(\ref{eq:eps}).},
\begin{align}\label{eq:symeps}
	\mathcal{E}(\pi\hbar n+q)=\mathcal{E}(\pi\hbar n-q),\quad 0\le q\le\pi\hbar n. 
\end{align}
Since $\mathcal{E}(p)$ is analytic in our case \cite{Note3}\footnotetext[3]{Note that this is not a generic situation since for certain models $\mathcal{E}(p)$  exhibits a cusp in some parameter regimes \cite{lamacraft_dispersion_2009,schecter_critical_2012}.}, the property (\ref{eq:symeps}) shows that odd derivatives of $\mathcal{E}(p)$ at its maximum, $p=\pi\hbar n$, nullify. 

For a given set of quasimomenta $\{k_1,k_2,\ldots,k_N,\eta\}$ that satisfies Eqs.~(\ref{BA}), there is another one that also satisfies Eqs.~(\ref{BA}). It is defined  by the shift
\begin{align}\label{eq:LLtr}
	\tilde k_j={}&k_j+\frac{2\pi}{L}\ell,\quad 1\le j\le N,\\
	\quad \tilde\eta={}&\eta+\frac{2\pi}{L}\ell,
\end{align} 
where $\ell$ is an integer. The energy and momentum of the new set are
\begin{subequations}
\begin{gather}
	\tilde E=\frac{\hbar^2}{2m}\sum_{j=1}^N \tilde k_j^2=E+\frac{2\pi\ell\;\! \hbar n}{Nm}\left(p+{\pi\ell\;\!\hbar n}\right),\\
	\tilde p=\hbar \sum_{j=1}^N \tilde k_j=p+2\pi \ell\;\! \hbar n,
\end{gather}	
\end{subequations}
which correspond to the energy and momentum (\ref{eq:EP}) of the original set. In the thermodynamic limit, the energies of the two configurations are the same, $\tilde E=E$, while the momentum is shifted by an integer multiple of $2\pi\hbar n$. Since the energy of the Lieb-Liniger model does not change \cite{lieb_exact_1963} under the transformation (\ref{eq:LLtr}), we conclude that the polaron energy is a periodic function of the momentum, $\mathcal{E}(p)=\mathcal{E}(p+2\pi\ell\;\!\hbar n)$.
  
The polaron energy (\ref{eq:eps}) is an even analytic function of the momentum (\ref{eq:p}). At low momenta it is therefore characterized by the Maclaurin series
\begin{align}\label{eq:E(p)}
	\mathcal{E}(p)=\frac{p^2}{2m^*}-\frac{\nu\;\! p^4}{24 \hbar^2 n^2 m}+\ldots,
\end{align}
which applies at arbitrary interaction and momenta $p\ll \hbar n \sqrt{{m}/{m^* \nu}}$. In Eq.~(\ref{eq:E(p)}), $\nu$ is the dimensionless parameter that controls the quartic term and is discussed later in the paper, while the quadratic term is controlled by
$m^*$, which denotes the mass of polaron excitation. It can be exactly expressed as \cite{ristivojevic_exact_2021}
\begin{align}\label{eq:m*LL}
	\frac{m}{m^*}=-\gamma^2 \frac{\partial}{\partial\gamma}\left(\frac{e(\gamma)}{\gamma^2}\right),
\end{align}
where the dimensionless function $e(\gamma)$ determines the ground-state energy per particle (\ref{eq:E0}) of the Lieb-Liniger model via \cite{lieb_exact_1963}
\begin{align}\label{eq:E0LL}
	\epsilon_0=\frac{\hbar^2 n^2}{2m}  e(\gamma).
\end{align}
Here $\gamma=c/n$ is the dimensionless parameter that accounts for the interaction strength between the bosons.

\section{Polaron excitation spectrum at weak interaction\label{sec3}}

In the regime of weak interaction, $\gamma\ll 1$, the integral Bethe ansatz Eq.~(\ref{eq:rho}) for a choice of the phase shift (\ref{eq:phaseshift}) was analyzed in Ref.~\cite{popov_theory_1977}. For $|k|<Q$, the solution at two leading orders is given by
\begin{align}\label{eq:rhosol}
	\rho(k,Q)=\frac{\sqrt{Q^2-k^2}}{2\pi c}+\frac{\left(1+\ln\frac{16\pi Q}{c}\right)Q-k\ln\frac{Q+k}{Q-k}}{4\pi^2 \sqrt{Q^2-k^2}}.
\end{align}
Substituting Eq.~(\ref{eq:rhosol}) in Eq.~(\ref{eq:n}) we obtain the density $n$ as a function of the Fermi rapidity $Q$. After the inversion one finds
\begin{align}\label{eq:Qsol}
	Q=2n\sqrt{\gamma}\left[1-\frac{\sqrt{\gamma}}{4\pi}\left(\ln\frac{32\pi}{\sqrt\gamma}-1\right)+O(\gamma)\right].
\end{align}
The analysis of Eq.~(\ref{eq:rho}) from Ref.~\cite{popov_theory_1977}  can be also applied to study Eq.~(\ref{eq:sigma}). At $|k|<Q$, at two leading orders we obtain
\begin{align}\label{eq:sigmasol}
	\sigma(k,Q{}&)=\frac{\hbar^2}{2m}\left[ \frac{k\sqrt{Q^2-k^2}}{c}\right.\notag\\
	 &\left.+\frac{\left(1+\ln\frac{16\pi Q}{c}\right)Qk-(2k^2-Q^2)\ln\frac{Q+k}{Q-k}}{2\pi \sqrt{Q^2-k^2}}\right].
\end{align}
We notice that both Eqs.~(\ref{eq:rhosol}) and (\ref{eq:sigmasol}) assume that $k$ is not in the close vicinity of $\pm Q$. The condition can be expressed as $1-k^2/Q^2\gg \sqrt{\gamma}$ \cite{popov_theory_1977}. Such inaccuracy is, however, not important for the accuracy of the results in this section.

Substituting Eqs.~(\ref{eq:rhosol}) and (\ref{eq:sigmasol}) into Eqs.~(\ref{eq:peps}), for $-Q\le \eta\le Q$ we find
\begin{subequations}
	\label{eq:epspweak}
\begin{align}
	\label{eq:pweak}
	p={}&\hbar n\left[2\phi-\sin(2\phi)+\sqrt\gamma A(\phi)+O(\gamma) \right]\\
	\mathcal{E}={}&\frac{4\hbar^2 n^2}{3 m}\sqrt{\gamma}\left[\sin^3\phi+ \frac{3\sqrt\gamma}{4}\left(A(\phi) \cos\phi-\frac{1}{2}\right)+O(\gamma)\right]
\end{align}
\end{subequations}
where
\begin{align}
	A(\phi)={}&\cos\phi +\frac{\sin(2\phi)}{2\pi}\left(\ln\frac{32\pi}{\sqrt\gamma}-1\right)\notag\\&
	+\frac{2\sin\phi}{\pi} \ln\tan\left(\frac{\phi}{2}\right).
\end{align}
Here we have introduced the parametrization $\cos\phi=\eta/Q$, where  $0\le\phi\le \pi$. The terms in brackets in Eqs.~(\ref{eq:epspweak}) proportional to $\sqrt\gamma$ arise from the subleading terms of Eqs.~(\ref{eq:rhosol}) and (\ref{eq:sigmasol}). They should, therefore, be smaller than the corresponding leading-order ones. At small momenta this occurs at
\begin{align}\label{eq:p*condition}
	p\gg p^*=\hbar n\sqrt{\gamma},
\end{align}
which is the condition for the applicability of the spectrum (\ref{eq:epspweak}). 

The evaluation of Eqs.~(\ref{eq:peps}) for $\eta\ge Q$ at the leading order in $\gamma\ll 1$ leads to
\begin{subequations}
	\label{eq:simple}
	\begin{align}
		\label{eq:pweaksmall}
		p={}&\hbar n\sqrt\gamma\;\! e^{-\phi},\\
		\mathcal{E}={}&\frac{\hbar^2 n^2}{2m}\gamma\;\! e^{-2\phi},
	\end{align}
\end{subequations} 
where $\eta=Q\cosh\phi$ and $\phi>0$. The form of Eq.~(\ref{eq:pweaksmall}) implies $p<p^*$, which is consistent with the condition (\ref{eq:p*condition}) obtained for the complementary region. Therefore, at momenta below $p^*$, we have obtained a quadratic spectrum of the polaron, $\mathcal{E}(p)=p^2/2m$,  which crosses over into the parametric form given by Eq.~(\ref{eq:epspweak}) at momenta above $p^*$. From the symmetry property of the spectrum given by Eq.~(\ref{eq:symeps}) follows that the spectrum is also quadratic in the vicinity of $2\pi\hbar n$. This can be also obtained by studying the case $\eta\le -Q$.

\begin{figure}
	\includegraphics[width=\columnwidth]{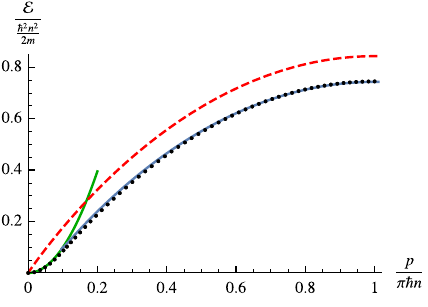}
	\caption{Comparison of the exact result for the polaron dispersion (the dots) and the obtained parametric form (solid curves) given by Eqs.~(\ref{eq:epspweak}) and (\ref{eq:simple}) for $\gamma=0.1$. The dashed curve is the dark soliton dispersion. On the plot is shown only the region of momenta $0\le p\le \pi\hbar n$, while for other $p$ the dispersion follows from the symmetry property (\ref{eq:symeps}) and the periodicity.} \label{fig1}
\end{figure}

In this section we have calculated the polaron spectrum (\ref{eq:epspweak}) at weak interaction, $\gamma\ll 1$. Interestingly, accounting only for the leading-order term (i.e., neglecting the terms proportional to $\sqrt{\gamma}$ in the brackets), Eqs.~(\ref{eq:epspweak}) describe the spectrum of the dark soliton solution  \cite{kulish_comparison_1976,ishikawa_solitons_1980,imambekov_one-dimensional_2012} of the Gross-Pitaevskii equation. It corresponds to type-II excitations in the Lieb-Liniger model \cite{lieb_exact_1963} beyond very small region of momenta of the characteristic size $\hbar n\gamma^{3/4}$ \cite{imambekov_one-dimensional_2012,pustilnik_low-energy_2014,ristivojevic_excitation_2014}. However, the polaron excitation energy is always smaller than the energy of type-II excitation with the same momentum, as we have explicitly shown in Eq.~(\ref{eq:ineq}) at the energy maximum, i.e., at $p=\pi\hbar n$. Equations (\ref{eq:epspweak}) at $\phi=\pi/2$ and thus $A(\phi)=0$ also illustrate this point, leading to 
\begin{align}\label{eq:top}
	\mathcal{E}(\pi\hbar n)=\frac{4\hbar^2 n^2}{3m}\sqrt\gamma\left(1-\frac{3\sqrt\gamma}{8}+O(\gamma)\right).
\end{align}
The leading-order term in Eq.~(\ref{eq:top}) is the energy of the dark soliton, while the whole expression (\ref{eq:top}) represents the polaron energy, which is smaller. Notice that the energy of the type-II excitation with the momentum $\pi\hbar n$ does not have the correction proportional to $\gamma$ \cite{reichert_exact_2019}, unlike the polaron. In Fig.~\ref{fig1} we show the exact result obtained numerically for the polaron dispersion and small-$\gamma$ expansion given by Eqs.~(\ref{eq:epspweak}) and (\ref{eq:simple}). The agreement is quite good even for not particularly small value $\gamma=0.1$, becoming better with decreasing $\gamma$. In the Appendix are given the details about the numerical procedure used to produce Fig.~\ref{fig1}.

\section{Polaron excitation spectrum at strong interaction\label{sec4}}

At strong interaction, $\gamma\gg 1$, the Bethe ansatz Eqs.~(\ref{eq:rho}) and (\ref{eq:sigma}) can be solved using a perturbative expansion controlled by $1/\gamma$ \cite{lieb_exact_1963,ristivojevic_excitation_2014}. Accounting for the first three orders, we find
\begin{gather}\label{eq:rhosignastrong}
	\rho(k,Q)=\frac{1}{2\pi}+ \frac{Q}{\pi^2 c}+\frac{2Q^2}{\pi^3 c^2}+O(Q^3/c^3),\\ \label{eq:sigmasignastrong}
	\sigma(k,Q)=\frac{\hbar^2 k}{m}+O(Q^3/c^3).
\end{gather}
This enables us to evaluate Eqs.~(\ref{eq:peps}), giving
\begin{subequations}\label{eq:pEstrong}
\begin{align}
	p={}&\pi\hbar n\left[1-\frac{2 \arctan\vartheta}{\pi}+\frac{8\pi\vartheta	}{3(1+\vartheta^2)^2 \gamma^2} +O(\gamma^{-3})\right],\\
	\mathcal{E}={}&\frac{\hbar^2 n^2}{2m}\biggl\{ \frac{8\pi^2}{3(1+\vartheta^2)\gamma}-\frac{16\pi^2}{(1+\vartheta^2) \gamma^2} +\biggl[\frac{64\pi^2}{1+\vartheta^2}\notag\\
	&+\frac{32\pi^4}{5(1+\vartheta^2)^2} -\frac{128\pi^4}{15(1+\vartheta^2)^3}\biggr]\frac{1}{\gamma^3}+O(\gamma^{-4})\biggr\},
\end{align}
\end{subequations}
where $\vartheta=2\eta/c$ is kept fixed.

\begin{figure}
	\includegraphics[width=\columnwidth]{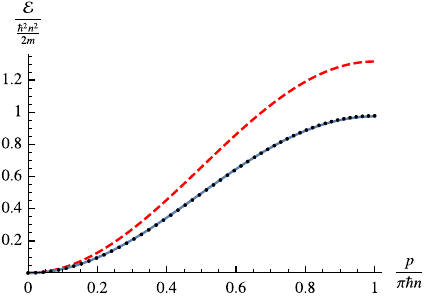}
	\caption{Comparison of the exact result for the polaron dispersion (the dots) and the obtained expression (\ref{eq:displarge}) (solid curve) for $\gamma=20$. The dashed curve represents the dispersion (\ref{eq:displarge}) at the leading order, i.e., accounting only for the $1/\gamma$ term.} \label{fig2}
\end{figure}

Equations (\ref{eq:pEstrong}) give the parametric form of the polaron excitation spectrum at strong interaction, where the parameter $\vartheta$ is a real number. We found that the spectrum (\ref{eq:pEstrong}) can be expressed explicitly in the form
\begin{subequations}\label{eq:displarge}
\begin{align}\label{eq:ansatz}
	\mathcal{E}(p)=\frac{\hbar^2 n^2}{2m} \sum_{j=1}^{+\infty} C_j(\gamma) \sin^{2j}\left(\frac{p}{2\hbar n}\right).
\end{align}
Substituting Eqs.~(\ref{eq:pEstrong}) in expression (\ref{eq:ansatz}) and evaluating it order by order in $1/\gamma$, we find the first three terms in the sum. They are given by
\begin{align}\label{eq:C1}
	C_1(\gamma)={}&\frac{8\pi^2}{3\gamma}-\frac{16\pi^2}{\gamma^2} +\frac{64\pi^2}{\gamma^3}+O(\gamma^{-4}),\\ \label{eq:C2}
	C_2(\gamma)={}&-\frac{32\pi^4}{45\gamma^3}+O(\gamma^{-4}),\\
	C_3(\gamma)={}&-\frac{64\pi^4}{45\gamma^3}+O(\gamma^{-4}).
\end{align}
\end{subequations}
We have verified that $C_4,C_5=O(\gamma^{-5})$. The coefficients $C_j$ therefore, very probably, decay at least as $\gamma^{-j}$, which makes the series (\ref{eq:ansatz}) rapidly converging. At the maximum, which occurs at $p=\pi\hbar n$, for the polaron energy we thus obtain
\begin{align}
	\mathcal{E}(\pi\hbar n)={}&\frac{4\pi^2\hbar^2 n^2}{3m\;\!\gamma}\biggr[1-\frac{6}{\gamma} +\frac{4(30-\pi^2)}{5\gamma^2}+O(\gamma^{-3})\biggl].
\end{align}
We notice that accounting for the leading-order term in Eq.~(\ref{eq:C1}), i.e., at $C_1=8\pi^2/3\gamma$ and thus taking $j=1$, dispersion (\ref{eq:displarge}) reduces to the result of Ref.~\cite{matveev_spectral_2008}. In Fig.~\ref{fig2} we compare the exact results for the dispersion with the analytical form (\ref{eq:displarge}). One can observe that even at moderately large $\gamma=20$, result (\ref{eq:displarge})  taken at the leading order shows significant deviation from the exact one, see Eq.~(\ref{eq:C1}). This occurs due to a relatively large ratio of the subleading and the leading terms in $C_1$. 

Ansatz  (\ref{eq:ansatz}) that remarkably simplifies the parametric dispersion (\ref{eq:pEstrong}) can be understood as a Fourier series of $\mathcal{E}(p)$ on the interval $[0,2\pi\hbar n]$ that satisfies the reflection property (\ref{eq:symeps}) and has even power series starting from $p^2$ around $p=0$. Expanding the form  (\ref{eq:ansatz}) at small $p$, one obtains the dispersion that coincides with Eq.~(\ref{eq:E(p)}) provided
\begin{gather}\label{eq:C1C2}
C_1=4\frac{m}{m^*},\quad C_2=\frac{4}{3}\left(\frac{m}{m^*}-\nu\right).
\end{gather}
We note that $m/m^*$ is given by Eq.~(\ref{eq:m*LL}) and $\nu$ is defined by Eq.~(\ref{eq:E(p)}). As a consistency check, we verified that the obtained value in Eq.~(\ref{eq:C1}) for $C_1$ is in full agreement with the general expression (\ref{eq:C1C2}) after substituting $m/m^*$ evaluated in Ref.~\cite{ristivojevic_exact_2021}. On the other hand, from  Eq.~(\ref{eq:C1C2}) we find $\nu=(C_1-3C_2)/4$. This yields
\begin{align}\label{eq:nugamma>>1}
	\nu=\frac{2\pi^2}{3\gamma}-\frac{4\pi^2}{\gamma^2}+\frac{8\pi^2(30+\pi^2)}{15\gamma^3}+O(\gamma^{-4})	
\end{align}
for the value of the other coefficient in Eq.~(\ref{eq:E(p)}) in the regime $\gamma\gg 1$.

The ansatz (\ref{eq:ansatz}) is not particularly useful in describing the polaron dispersion in the regime of weak interaction, $\gamma\ll 1$. Comparing the dispersion (\ref{eq:dispersionperturbationtheory}) that is derived further below with Eq.~(\ref{eq:ansatz}), we obtain
\begin{gather}
	\label{eq:C1C2weak}
	C_1=4+O(\sqrt\gamma),\quad C_2=-\frac{32}{5\pi\sqrt\gamma}+O(1),\notag\\
	C_3=-\frac{128}{7\pi\gamma^{3/2}}+O(1/\sqrt\gamma).
\end{gather} 
Therefore the series (\ref{eq:ansatz}) will be slowly converging at $p\sim \hbar n$ and $\gamma\ll 1$ and it appears that we need infinitely many terms to accurately describe the dispersion, in striking contrast to very few in the case $\gamma\gg 1$.

A motivated reader can follow the procedure of this section in order to evaluate further terms of the expansion in $1/\gamma$ in the spectrum (\ref{eq:displarge}). For this one needs solutions (\ref{eq:rhosignastrong}) and (\ref{eq:sigmasignastrong}) of the Bethe ansatz Eqs.~(\ref{eq:rho}) and (\ref{eq:sigma}) to higher orders, which can be obtained using a systematic method developed in Ref.~\cite{ristivojevic_excitation_2014}. We notice that the low momentum expansion of the dispersion (\ref{eq:ansatz}) will have a form similar to Eq.~(\ref{eq:E(p)}), where higher order even powers of $p$ should be included. Therefore, the knowledge of the Maclauren series of the dispersion at strong interaction is sufficient to infer the coefficients $C_j$ of Eq.~(\ref{eq:ansatz}). Vice versa is trivially correct since ansatz (\ref{eq:ansatz}) does not have restrictions on $p$.

\section{Discussion \label{sec5}}

The polaron dispersion at weak interactions can be calculated in a more elementary way. Consider a weakly interacting one-dimensional Bose gas of $N$ particles with the mass $m$ and the local repulsion $g$. Such a system is described by the Lieb-Liniger model (\ref{eq:H}) with the dimensionless parameter $\gamma=mg/\hbar^2 n$, where the density is $n=N/L$. We study an impurity of the mass $M$ and the momentum $P$ coupled locally to the Bose gas, where $G$ is the coupling strength. The correction to the energy of a slow impurity follows from the second-order  perturbation theory. The actual calculation is similar to the one performed in Appendix D of Ref.~\cite{reichert_quasiparticle_2017}, where we need  results (D1) and (D15) evaluated for $k_F=0$. They determine the energy of our system at the leading non-trivial order, 
\begin{subequations}\label{eq:dispersionperturbationtheo}
\begin{align}\label{eq:50a}
	E(P)=N\epsilon_0+G n+\frac{P^2}{2M}-\frac{Mv^2}{K } \frac{G^2}{g^2}\frac{\textrm{arctanh}\Delta(P)}{\Delta(P)}, 
\end{align}
where
	\begin{align}
		\Delta(P)={}&\sqrt{1-\frac{M^2}{m^2}+\frac{P^2}{m^2 v^2}}.	
	\end{align}
\end{subequations}
In Eq.~(\ref{eq:50a}), $N\epsilon_0$ denotes the ground-state energy of the isolated Bose gas [cf.~Eq.~(\ref{eq:E0LL})], while the constant term $Gn$ is the energy correction due to the presence of the impurity. In Eq.~(\ref{eq:50a}), $K$ is the Luttinger liquid parameter, which at weak interaction, $\gamma\ll 1$, is given by $K=\pi/\sqrt\gamma$, and $v=\pi \hbar n/mK$ is the sound velocity. The polaron dispersion is $\mathcal{E}(P)=E(P)-E(0)$ and reads
\begin{align}\label{eq:dispersionperturbationtheory}
		\mathcal{E}(P)={}&\frac{P^2}{2M}-\frac{Mv^2}{K} \frac{G^2}{g^2}\left(\frac{\textrm{arctanh}\Delta(P)}{\Delta(P)} - \frac{\textrm{arctanh}\Delta(0)}{\Delta(0)}\right).
\end{align}
We have assumed that $G\ll g\sqrt{K}$, since the correction to the bare spectrum $P^2/2M$ should be small.

In the special case $m=M$ and $g=G$, the dispersion (\ref{eq:dispersionperturbationtheory}) should correspond to the one of a polaron in the Yang-Gaudin model. Indeed, Eq.~(\ref{eq:dispersionperturbationtheory}) in this case reduces to the form (\ref{eq:E(p)}) with
\begin{align}\label{fin}
\frac{m}{m^*}=1-\frac{2}{3\pi}\sqrt\gamma,\quad \nu=\frac{24}{5\pi\sqrt\gamma},
\end{align} 
in agreement with Refs.~\cite{fuchs_spin_2005,ristivojevic_exact_2021}. On the other hand Eq.~(\ref{eq:dispersionperturbationtheory}) is more general as it is not restricted to the case of equal masses and coupling constants. Moreover, Eq.~(\ref{eq:dispersionperturbationtheory}) contains the leading-order result for the polaron dispersion at all higher powers of the momentum. The special case $m=M$ and $g=G$ of the latter can, in principle, be obtained from the Bethe ansatz, but this way is more complicated.

Our result (\ref{eq:dispersionperturbationtheo}) also contains the information about the binding energy of a polaron, which is the difference of the ground-state energies with and without it. This leads to 
\begin{align}
	\mu=Gn\left(1-\frac{\sqrt{\gamma}}{\pi}\frac{MG}{mg} \frac{\textrm{arctanh} \sqrt{1-\frac{M^2}{m^2}}}{ \sqrt{1-\frac{M^2}{m^2}}} \right).
\end{align}
In the special case $m=M$ and $g=G$ the latter result reduces to $\mu=(\hbar^2 n^2\gamma/m)(1-\sqrt\gamma/\pi)$, which is the polaron binding energy in the Yang-Gaudin model in the regime of weak interaction. It corresponds to the chemical potential of the Bose gas \cite{lieb_exact2_1963}, $\mu=\partial (N\epsilon_0)/\partial N$, as one can easily verify using Eq.~(\ref{eq:E0LL}) with $e(\gamma)=\gamma(1-4\sqrt\gamma/3\pi)$. We note that the energy (\ref{eq:50a}) can be actually expressed as $E(P)=N\epsilon_0+\mu+\mathcal{E}(P)$.

The results derived in this paper have a direct application. The polaron dispersion denotes the lower spectral edge for zero-temperature correlation functions, e.g., the dynamic  structure factor and the spectral function \cite{matveev_spectral_2008,kamenev_dynamics_2009,zvonarev_edge_2009,imambekov_one-dimensional_2012}. The latter are characterized by power-law singularities at the edge. The corresponding exponents are, quite generally, expressed in terms of the sound velocity and the corresponding Luttinger liquid parameter, as well as the derivatives of the dispersion with respect to the momentum and the density  \cite{kamenev_dynamics_2009,imambekov_one-dimensional_2012}. The latter directly follows from the results of this paper. Study of exponents of the correlation functions is left for a future work.

\appendix
\section{Numerical evaluation from Bethe ansatz \label{appendixnum}}

Here we give the detailed procedure for the numerical evaluation of the polaron dispersion. The numerical procedure is elementary, yet it can lead to highly precise results, if needed. It is based on the approach of Ref.~\cite{ristivojevic_conjectures_2019} developed to study the ground-state energy of the Lieb-Liniger model, which is here extended to account for the specific form of Eqs.~(\ref{eq:peps}). We begin by introducing the dimensionless parameters and functions,
\begin{gather}
\lambda=c/Q,\quad \beta=\eta/Q,\\
\varrho(x,\lambda)=\rho(Q x,Q),\quad \varsigma(x,\lambda)=\frac{m}{\hbar^2Q}\sigma(xQ,Q).
\end{gather}
Then the set of Bethe ansatz Eqs.~(\ref{eq:rho}) and (\ref{eq:sigma}), respectively, become
\begin{subequations}\label{eq:BAdimensionless}
\begin{gather}\label{eq1}
	\varrho(x,\lambda)-\frac{\lambda}{\pi}\int_{-1}^1dy\frac{\varrho(y,\lambda)}{\lambda^2+(x-y)^2} =\frac{1}{2\pi},\\
	\label{eq2}
	\varsigma(x,\lambda)-\frac{\lambda}{\pi}\int_{-1}^1dy\frac{\varsigma(y,\lambda)}{\lambda^2+(x-y)^2} = x,
\end{gather}
\end{subequations}
while from Eq.~(\ref{eq:n}) one finds
\begin{align}\label{eq3}
\frac{\lambda}{\gamma}=\int_{-1}^1dx\varrho(x,\lambda).
\end{align}
The momentum (\ref{eq:p}) and the energy (\ref{eq:eps}) of the polaron excitation are given by
\begin{subequations}\label{eq:pEdimensionless}
\begin{gather}
	p=\pi\hbar n\left[1+\frac{2\gamma}{\pi \lambda} \int_{-1}^{1} dx \varrho(x,\lambda)\arctan\left(\frac{2x-2\beta}{\lambda}\right)\right],\\
	\mathcal{E}=\frac{\hbar^2 n^2\gamma^2}{\pi m\lambda^2} \int_{-1}^{1} dx \varsigma(x,\lambda)\arctan\left(\frac{2x-2\beta}{\lambda}\right).
\end{gather}
\end{subequations}

We can solve Eqs.~(\ref{eq:BAdimensionless}) using the procedure of Ref.~\cite{ristivojevic_conjectures_2019} developed to solve Eq.~(\ref{eq1}). We adopt it here to study Eq.~(\ref{eq2}). We seek the solutions in the forms
\begin{subequations}\label{eq:varrhovarsigma}
\begin{gather}
	\varrho(x,\lambda)=\sum_{j=0}^{M} c_j(\lambda)T_{2j}(x),\\
	\varsigma(x,\lambda)=\sum_{j=0}^{M} d_j(\lambda)T_{2j+1}(x).
\end{gather}
\end{subequations}
Here $T_j(x)$ are the Chebyshev polynomials of the first kind, $T_n(x)=\cos(n \arccos x)$, and thus the assumed form takes into account the parity of $\varrho$ and $\varsigma$. The positive integer $M$ is arbitrary and should be adjusted to achieve the wanted precision in the final results. 

Substituting Eqs.~(\ref{eq:varrhovarsigma}) in Eqs.~(\ref{eq:BAdimensionless}) and (\ref{eq:pEdimensionless}), one obtains the integrals of the forms
\begin{subequations}\label{eq:FG}
	\begin{gather}
		F_j(x,\lambda)=\frac{\lambda}{\pi}\int_{-1}^{1}d y \frac{T_j(y)}{\lambda^2+(x-y)^2},\\
		A_j(\beta,\lambda)=\int_{-1}^{1}d y\:\! T_j(y)\arctan\left(\frac{2y}{\lambda}-\frac{2\beta}{\lambda}\right).
	\end{gather}
\end{subequations}
They can be evaluated with the help of the recurrent relations for the Chebyshev polynomials, $T_{j+1}(x)=2x T_j(x)-T_{j-1}(x)$ for integer $j\geq 1$. Introducing auxiliary integrals
\begin{gather}\label{eq:definitionG}
G_j(x,\lambda)=\frac{\lambda}{\pi}\int_{-1}^{1}d y \frac{2y\,T_j(y)}{\lambda^2+(x-y)^2},
\end{gather}
we find the recurrent relations
\begin{subequations}\label{eq:FGrecurence}
	\begin{align}
		F_j(x,\lambda)={}&G_{j-1}(x,\lambda)-F_{j-2}(x,\lambda),\\
		G_j(x,\lambda)={}&-\frac{4\lambda}{\pi} \frac{1-(-1)^j}{j(j-2)} +4x G_{j-1}(x,\lambda)-G_{j-2}(x,\lambda)\notag\\
		&-4(x^2+\lambda^2)F_{j-1}(x,\lambda),
	\end{align}
\end{subequations}
for $j\geq 2$. At $j=2$, the term $\frac{1-(-1)^j}{j(j-2)}$ must be set to zero. The functions $F_j$, and $G_j$  and at $j=0,1$ can be found directly from their definition, while for $j\geq2$ they can conveniently be calculated using Eqs.~(\ref{eq:FGrecurence}). The remaining function $A_j$ at $j=0,1$ can be found from its definition, while
\begin{align}\label{eq:Arecurrence}
A_j(\beta,\lambda)={}&\frac{\arctan\left(\frac{2\beta-2}{\lambda}\right)+(-1)^j \arctan\left(\frac{2\beta+2}{\lambda}\right)}{j^2-1}\notag\\
&-\frac{\pi F_{j+1}(\beta,\lambda/2)}{2(j+1)}	+\frac{\pi F_{j-1}(\beta,\lambda/2)}{2(j-1)}	
\end{align}
for $j\geq 2$. Equations~(\ref{eq:BAdimensionless}) then become
\begin{subequations}\label{eq:A11}
\begin{gather}\label{eq:clambda}
	\sum_{j=0}^M c_j(\lambda)\left[T_{2j}(x)-F_{2j}(x,\lambda)\right]=\frac{1}{2\pi},\\
	\sum_{j=0}^M d_j(\lambda)\left[T_{2j+1}(x)-F_{2j+1}(x,\lambda)\right]=x,
	\end{gather}
\end{subequations}
while the condition (\ref{eq3}) determines the Lieb parameter,
\begin{align}\label{eq:gammafinal}
	\gamma=\frac{\lambda}{\sum_{j=0}^M \frac{2c_j(\lambda)}{1-4j^2}}.
\end{align}
Finally, from Eqs.~(\ref{eq:pEdimensionless}) we find
\begin{subequations}\label{eq:A13}
\begin{gather}
	p=\pi\hbar n \left[1+\frac{2\gamma}{\pi\lambda}\,{\sum_{j=0}^M c_j(\lambda) A_{2j}(\beta,\lambda)}\right],\\
	\mathcal{E}=\frac{\hbar^2 n^2\gamma^2}{\pi m\lambda^2}\, {\sum_{j=0}^M d_j(\lambda) A_{2j+1}(\beta,\lambda)}.
\end{gather}	
\end{subequations}
In Eqs.~(\ref{eq:A13}) one should express $\gamma$ from Eq.~(\ref{eq:gammafinal}).

The application of the previous recipe for the calculation of the polaron dispersion is simple. For a given $\lambda>0$ and the upper limit of summation $M$ one should generate the sequences $A_j$, $F_j$, and $G_j$ for $j=0,1,\ldots, 2M+1$. For $j=0,1$ one evaluates the integrals from the definitions (\ref{eq:FG}) and (\ref{eq:definitionG}), while at $j\ge 2$ it is convenient to use the derived recurrent relations (\ref{eq:FGrecurence}) and (\ref{eq:Arecurrence}). This enables one to solve the linear Eqs.~(\ref{eq:A11}). One finally evaluates $\gamma$, the polaron momentum and energy using Eqs.~(\ref{eq:gammafinal}) and (\ref{eq:A13}). A particular feature of the present algorithm is that is enables one to obtain numerical results at huge precision that is systematically increased when increasing $M$. This is possible since  $A_j$, $F_j$, and $G_j$ are evaluated analytically and thus can be obtained numerically at any needed precision, enabling one to solve the linear Eqs.~(\ref{eq:A11}) at high precision. The efficiency of the algorithm for the Lieb-Liniger case was discussed in Ref.~\cite{ristivojevic_conjectures_2019}, while the present one has similar features.


%

\end{document}